\documentclass[nonacm,sigplan]{acmart}
\usepackage{subfig}
\usepackage{graphicx}
\usepackage{subcaption}

\begin{document}

\title{Towards Forever Access for Implanted Brain-Computer Interfaces}

\author{Muhammed Ugur}
%\email{muhammed.ugur@yale.edu}
\affiliation{
  \institution{Yale University}
  \postcode{06511}
  \country{}
}
\author{\centerline{\mbox{Raghavendra Pradyumna Pothukuchi}}}
%\email{}
\affiliation{
  \institution{Yale University}
  \postcode{06511}
  \country{}
}
\author{Abhishek Bhattacharjee}
%\email{abhishek.bhattacharjee@yale.edu}
\affiliation{
  \institution{Yale University}
  \postcode{06511}
  \country{}
}

\vspace{2mm}

\begin{abstract}
Designs for implanted brain-computer interfaces (BCIs) have increased significantly in recent years. Each device promises better clinical outcomes and quality-of-life improvements, yet due to severe and inflexible safety constraints, progress requires tight co-design from materials to circuits and all the way up the stack to applications and algorithms. This trend has become more aggressive over time, forcing clinicians and patients to rely on vendor-specific hardware and software for deployment, maintenance, upgrades, and replacement. This over-reliance is ethically problematic, especially if companies go out-of-business or business objectives diverge from clinical promises. Device heterogeneity additionally burdens clinicians and healthcare facilities, adding complexity and costs for in-clinic visits, monitoring, and continuous access. 

Reliability, interoperability, portability, and future-proofed design is needed, but this unfortunately comes at a cost. These system features sap resources that would have otherwise been allocated to reduce power/energy and improve performance. Navigating this trade-off in a systematic way is critical to providing patients with {\it forever access} to their implants and reducing burdens placed on healthcare providers and caretakers. We study the integration of on-device storage to highlight the sensitivity of this trade-off and establish other points of interest within BCI design that require careful investigation. In the process, we revisit relevant problems in computer architecture and medical devices from the current era of hardware specialization and modern neurotechnology.

\end{abstract}

\maketitle 
\pagestyle{plain}

\section{Introduction}

Brain-computer interfaces (BCIs) read the biological activity of neurons and provide treatment through electrical stimulation, send commands to external devices, or improve our understanding of the brain and its circuitry. The most effective BCIs are surgically implanted, providing two orders of magnitude better signal quality then non-implanted approaches \cite{SNR2009}. Implanted BCIs require the placement of microelectrodes and probes in, or on the surface of the brain. These probes can have multiple channels (i.e., sensors for neural activity), and the signals from them are digitized and processed.

\begin{figure}[h]
\centering
\includegraphics[width=0.50\textwidth]{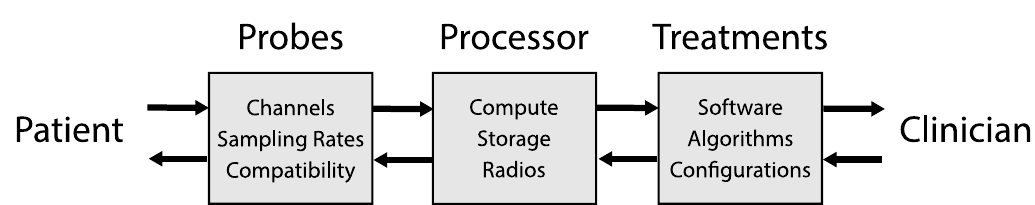}
\caption{High level components in an implanted BCI, and natural system boundaries to explore interoperability mechanisms. Each component can has several sub-components (e.g., the compute and storage in the processor), or various features (e.g., the sampling rate of the probes), which all play a role in determining interoperability.}
\label{fig:boundaries}
\vspace{-4mm}
\end{figure}

Providing continuous access to neural interfaces after implantation is becoming an increasingly serious ethical concern as BCI use becomes more widespread. There have been reports of companies, both in clinical trials and after regulatory approval, abandoning patients who have had their devices implanted, due to bankruptcy and inadequate post-trial planning and support \cite{Drew2020, BionicEye2022, Abandoned2022, HamzelouExplant2023}. These instances are alarming and have led to patients abruptly losing access to life-changing therapies and treatments. This has even led to coercive explantation procedures. Explantation is when the device and its electrodes have to be completely removed from the patient's body, both for the patient's safety and also to eliminate legal and financial risks for stakeholders \cite{HamzelouExplant2023}. 

Explantation is a consequence of poor interoperability. For FDA-approval, companies typically design an implant with a single algorithm to treat a single disease, performing tight co-design from the probes to the applications. This is because the approval process is very disease- and treatment-specific and requires adhering to strict power and safety constraints. This, plus the proprietary nature of BCI development, discourages the added cost of interoperability and other levels of flexibility. Recent work has encouraged the shift towards flexibility through on-chip reconfigurable pipelines of accelerators to support many different applications \cite{karageorgos2020hardware, Sriram2023}.

Aggressive co-design, however, still requires working with different teams, levels of expertise, and intellectual property to design an implant that is safe, reliable, and correct. This has introduced natural boundaries to specify interoperability. Figure \ref{fig:boundaries} shows the typical components of a BCI and its boundaries. Naively using these boundaries for interoperability is not enough. Providing effective treatment and better quality-of-life requires intelligent co-design of specifications across these boundaries. For example, individuals may require different channel count and sampling rate configurations based on diagnosis. This personalization at the probe-level can stress on-device compute and algorithms \cite{Swapping2024}. Channels may have to be ignored or sampling rates may need to be lowered to handle real-time processing. \textit{We argue that principled abstractions and system layers are needed to utilize these components for the best possible treatment and quality of care.} 

\section{Background}
In recent years, the number of simultaneously recorded neurons have increased exponentially \cite{Stevenson2011}. This, coupled with better quality-of-life features like wireless transmission and recharging, has stressed the power constraints of implanted BCIs. This has led to more intricate control logic and custom chips which support on-device compute, storage, and radios to manage this growing data volume. 

As devices become more complex, maintenance costs increase. These costs encompass clinical visits, infection management, lead revision, battery replacement, device repair, software monitoring, technical support, and more \cite{BaitAndSwitch2017, SurgeryCosts2017, ContinuedAccess2018}. One or more of these costs range from \$10,000-\$100,000 and may be prohibitively expensive for researchers, companies, and healthcare providers to support many years into the future, especially as the number of patients and use cases increase. 

These costs become heavier as different companies produce multiple generations of proprietary devices. The Breakthrough Devices Program by the FDA \cite{BreakthroughDevices} and recent successes in clinical trials have expanded device heterogeneity. This creates cross-platform and compatibility issues when problems arise, e.g., when companies go out-of-business, healthcare facilities lack device support, physicians recommend updates for the patient's evolving condition, replacements become too expensive so cheaper alternatives are used, and so on. Interoperability is becoming a necessity for BCI design to avoid putting patients through risky procedures and terminating/downgrading their treatment. 

\section{Processor-Probe Compatibility}

\begin{figure}[h]
\centering
\includegraphics[width=0.45\textwidth]{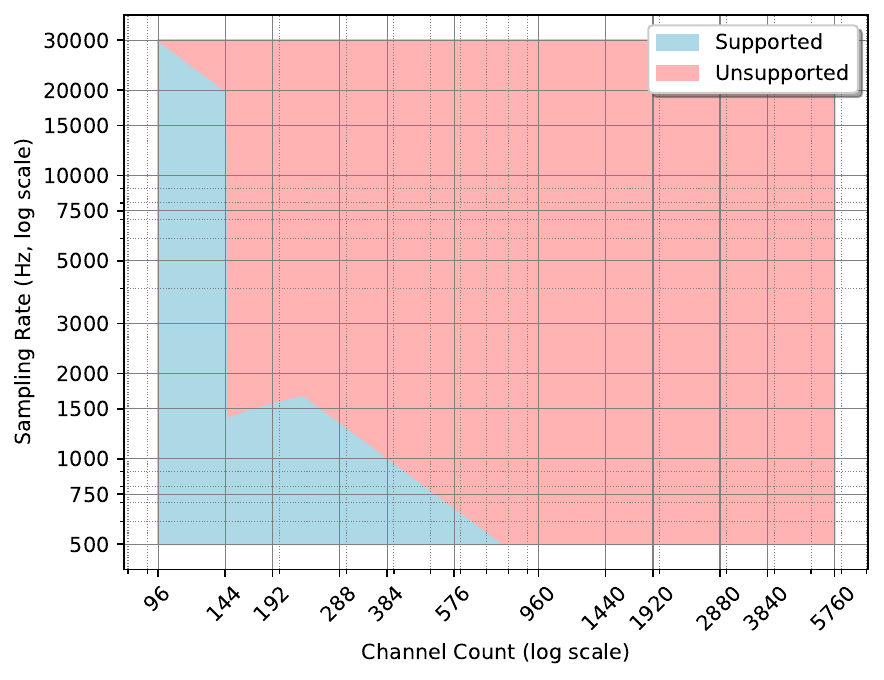}
\caption{Shows the support of channel/sampling-rate configurations (blue region) using on-device storage and naive swapping for a signal processing accelerator \cite{Swapping2024}.}
\label{fig:channels}
\end{figure}

One approach to add interoperability between processors and probes is through intelligently using storage, which serves as a swap space to support the data ingestion from a flexible number of probes. Storage is already being used in BCIs to store raw electrophysiology data over periods of time for querying and periodic monitoring. In principle, its smart integration could help alleviate the pressure on computation power for higher channel counts and data rates. However, doing so naively will limit configuration support. Figure \ref{fig:channels} shows how a naive swapping approach can be used to improve interoperability between processors and probes \cite{Swapping2024}. This example shows how storage gives some flexibility but is ultimately limited by bandwidth, latency, and accelerator design. Navigating this trade-off is crucial and has significant implications in personalized treatment and care.

Conventional computer systems have embraced this trade-off between performance and flexibility in the form of virtual memory. Historically, virtual memory was introduced as a performance tax for software portability, resource multiplexing, and ease of programmability. It leveraged memory persistence and capacity to provide these features, resembling the approach from Figure \ref{fig:channels}. The distinction with BCIs is that flexibility comes at the cost of power, which is already severely limited, and takes away resources that could otherwise improve the device's primary function. This requires a re-envisioning of virtual memory-like principles from the perspective of power and hardware specialization. The ultimate choice of abstraction and layering for flexibility will have significant impact on patient health and quality-of-life. The overhead of interoperability must be low and requires systematic study. Luckily, many of the algorithms are known and predictable, providing abundant room for improvement compared to general-purpose approaches.

\section{Previous Work}
Previous work has outlined post-trial responsibilities and encouraged the availability and compatibility of replacement hardware and software \cite{Hendriks2019}. Past efforts in cardiac pacemakers provide examples for how standardization and cross-compatibility between parts can be achieved \cite{PacemakerEthics2016, ISO-5841-3:2013}. However, BCIs pose additional challenges. Their use cases span many different diseases, from movement disorders to epilepsy, and require strict safety measures given their nascency and proximity to the brain. 

\section{Next Steps}
Navigating interoperability for invasive neural interfaces requires regulatory and technical guidance from all relevant stakeholders. Architects have experience  balancing this performance-flexibility trade-off and should work closely with clinicians, policymakers, and patients to layout a plan for adoption and implementation. The end result will have direct impact on the quality-of-life and treatment of patients suffering from severe neurological and psychiatric disorders.

\bibliographystyle{ACM-Reference-Format}
\bibliography{references}

%%% -*-BibTeX-*-
%%% Do NOT edit. File created by BibTeX with style
%%% ACM-Reference-Format-Journals [18-Jan-2012].

\begin{thebibliography}{16}

%%% ====================================================================
%%% NOTE TO THE USER: you can override these defaults by providing
%%% customized versions of any of these macros before the \bibliography
%%% command.  Each of them MUST provide its own final punctuation,
%%% except for \shownote{}, \showDOI{}, and \showURL{}.  The latter two
%%% do not use final punctuation, in order to avoid confusing it with
%%% the Web address.
%%%
%%% To suppress output of a particular field, define its macro to expand
%%% to an empty string, or better, \unskip, like this:
%%%
%%% \newcommand{\showDOI}[1]{\unskip}   % LaTeX syntax
%%%
%%% \def \showDOI #1{\unskip}           % plain TeX syntax
%%%
%%% ====================================================================

\ifx \showCODEN    \undefined \def \showCODEN     #1{\unskip}     \fi
\ifx \showDOI      \undefined \def \showDOI       #1{#1}\fi
\ifx \showISBNx    \undefined \def \showISBNx     #1{\unskip}     \fi
\ifx \showISBNxiii \undefined \def \showISBNxiii  #1{\unskip}     \fi
\ifx \showISSN     \undefined \def \showISSN      #1{\unskip}     \fi
\ifx \showLCCN     \undefined \def \showLCCN      #1{\unskip}     \fi
\ifx \shownote     \undefined \def \shownote      #1{#1}          \fi
\ifx \showarticletitle \undefined \def \showarticletitle #1{#1}   \fi
\ifx \showURL      \undefined \def \showURL       {\relax}        \fi
% The following commands are used for tagged output and should be
% invisible to TeX
\providecommand\bibfield[2]{#2}
\providecommand\bibinfo[2]{#2}
\providecommand\natexlab[1]{#1}
\providecommand\showeprint[2][]{arXiv:#2}

\bibitem[ISO(2013)]%
        {ISO-5841-3:2013}
 \bibinfo{year}{2013}\natexlab{}.
\newblock \bibinfo{booktitle}{\emph{ISO 5841-3:2013}}.
\newblock
\urldef\tempurl%
\url{https://www.iso.org/standard/60542.html}
\showURL{%
Retrieved March 25, 2023 from \tempurl}


\bibitem[Ball et~al\mbox{.}(2009)]%
        {SNR2009}
\bibfield{author}{\bibinfo{person}{Tonio Ball}, \bibinfo{person}{Markus Kern}, \bibinfo{person}{Isabella Mutschler}, \bibinfo{person}{Ad Aertsen}, {and} \bibinfo{person}{Andreas Schulze-Bonhage}.} \bibinfo{year}{2009}\natexlab{}.
\newblock \showarticletitle{Signal quality of simultaneously recorded invasive and non-invasive EEG}.
\newblock \bibinfo{journal}{\emph{NeuroImage}} \bibinfo{volume}{46}, \bibinfo{number}{3} (\bibinfo{date}{July} \bibinfo{year}{2009}), \bibinfo{pages}{708–716}.
\newblock
\showISSN{1053-8119}
\urldef\tempurl%
\url{https://doi.org/10.1016/j.neuroimage.2009.02.028}
\showDOI{\tempurl}


\bibitem[Chen et~al\mbox{.}(2017)]%
        {SurgeryCosts2017}
\bibfield{author}{\bibinfo{person}{Tsinsue Chen}, \bibinfo{person}{Zaman Mirzadeh}, \bibinfo{person}{Margaret Lambert}, \bibinfo{person}{Omar Gonzalez}, \bibinfo{person}{Ana Moran}, \bibinfo{person}{Andrew~G. Shetter}, {and} \bibinfo{person}{Francisco~A. Ponce}.} \bibinfo{year}{2017}\natexlab{}.
\newblock \showarticletitle{Cost of Deep Brain Stimulation Infection Resulting in Explantation}.
\newblock \bibinfo{journal}{\emph{Stereotactic and Functional Neurosurgery}} \bibinfo{volume}{95}, \bibinfo{number}{2} (\bibinfo{year}{2017}), \bibinfo{pages}{117–124}.
\newblock
\showISSN{1423-0372}
\urldef\tempurl%
\url{https://doi.org/10.1159/000457964}
\showDOI{\tempurl}


\bibitem[Drew(2020)]%
        {Drew2020}
\bibfield{author}{\bibinfo{person}{Liam Drew}.} \bibinfo{year}{2020}\natexlab{}.
\newblock \showarticletitle{“Like taking away a part of myself” ― life after a neural implant trial}.
\newblock \bibinfo{journal}{\emph{Nature Medicine}} \bibinfo{volume}{26}, \bibinfo{number}{8} (\bibinfo{date}{July} \bibinfo{year}{2020}), \bibinfo{pages}{1154–1156}.
\newblock
\showISSN{1546-170X}
\urldef\tempurl%
\url{https://doi.org/10.1038/d41591-020-00028-8}
\showDOI{\tempurl}


\bibitem[Drew(2022)]%
        {Abandoned2022}
\bibfield{author}{\bibinfo{person}{Liam Drew}.} \bibinfo{year}{2022}\natexlab{}.
\newblock \bibinfo{booktitle}{\emph{Abandoned: The human cost of neurotechnology failure}}.
\newblock
\urldef\tempurl%
\url{https://www.nature.com/immersive/d41586-022-03810-5/index.html}
\showURL{%
Retrieved March 25, 2023 from \tempurl}


\bibitem[Eliza~Strickland(2022)]%
        {BionicEye2022}
\bibfield{author}{\bibinfo{person}{Mark~Harris Eliza~Strickland}.} \bibinfo{year}{2022}\natexlab{}.
\newblock \bibinfo{booktitle}{\emph{IEEE Spectrum: Their Bionic Eyes are Now Obsolete and Unsupported}}.
\newblock
\urldef\tempurl%
\url{https://spectrum.ieee.org/bionic-eye-obsolete}
\showURL{%
Retrieved March 25, 2023 from \tempurl}


\bibitem[et. al.(2019)]%
        {Hendriks2019}
\bibfield{author}{\bibinfo{person}{Hendriks et. al.}} \bibinfo{year}{2019}\natexlab{}.
\newblock \showarticletitle{Ethical Challenges of Risk, Informed Consent, and Posttrial Responsibilities in Human Research With Neural Devices: A Review}.
\newblock \bibinfo{journal}{\emph{JAMA Neurology}} \bibinfo{volume}{76}, \bibinfo{number}{12} (\bibinfo{date}{Dec.} \bibinfo{year}{2019}), \bibinfo{pages}{1506}.
\newblock
\showISSN{2168-6149}
\urldef\tempurl%
\url{https://doi.org/10.1001/jamaneurol.2019.3523}
\showDOI{\tempurl}


\bibitem[Food and (FDA)(2024)]%
        {BreakthroughDevices}
\bibfield{author}{\bibinfo{person}{U.S. Food} {and} \bibinfo{person}{Drug~Administration (FDA)}.} \bibinfo{year}{2024}\natexlab{}.
\newblock \bibinfo{booktitle}{\emph{Breakthrough Devices Program}}.
\newblock
\urldef\tempurl%
\url{https://www.fda.gov/medical-devices/how-study-and-market-your-device/breakthrough-devices-program}
\showURL{%
Retrieved March 25, 2023 from \tempurl}


\bibitem[Hamzelou(2023)]%
        {HamzelouExplant2023}
\bibfield{author}{\bibinfo{person}{Jessica Hamzelou}.} \bibinfo{year}{2023}\natexlab{}.
\newblock \bibinfo{booktitle}{\emph{MIT Technology Review: A brain implant changed her life. Then it was removed against her will.}}
\newblock
\urldef\tempurl%
\url{https://www.technologyreview.com/2023/05/25/1073634/brain-implant-removed-against-her-will/}
\showURL{%
Retrieved March 25, 2023 from \tempurl}


\bibitem[Hutchison and Sparrow(2016)]%
        {PacemakerEthics2016}
\bibfield{author}{\bibinfo{person}{Katrina Hutchison} {and} \bibinfo{person}{Robert Sparrow}.} \bibinfo{year}{2016}\natexlab{}.
\newblock \showarticletitle{What Pacemakers Can Teach Us about the Ethics of Maintaining Artificial Organs}.
\newblock \bibinfo{journal}{\emph{Hastings Center Report}} \bibinfo{volume}{46}, \bibinfo{number}{6} (\bibinfo{date}{Nov.} \bibinfo{year}{2016}), \bibinfo{pages}{14–24}.
\newblock
\showISSN{1552-146X}
\urldef\tempurl%
\url{https://doi.org/10.1002/hast.644}
\showDOI{\tempurl}


\bibitem[Karageorgos et~al\mbox{.}(2020)]%
        {karageorgos2020hardware}
\bibfield{author}{\bibinfo{person}{Ioannis Karageorgos}, \bibinfo{person}{Karthik Sriram}, \bibinfo{person}{J{\'a}n Vesel{\`y}}, \bibinfo{person}{Michael Wu}, \bibinfo{person}{Marc Powell}, \bibinfo{person}{David Borton}, \bibinfo{person}{Rajit Manohar}, {and} \bibinfo{person}{Abhishek Bhattacharjee}.} \bibinfo{year}{2020}\natexlab{}.
\newblock \showarticletitle{Hardware-software co-design for brain-computer interfaces}. In \bibinfo{booktitle}{\emph{2020 ACM/IEEE 47th Annual International Symposium on Computer Architecture (ISCA)}}. IEEE, \bibinfo{pages}{391--404}.
\newblock


\bibitem[Lázaro-Muñoz et~al\mbox{.}(2018)]%
        {ContinuedAccess2018}
\bibfield{author}{\bibinfo{person}{Gabriel Lázaro-Muñoz}, \bibinfo{person}{Daniel Yoshor}, \bibinfo{person}{Michael~S. Beauchamp}, \bibinfo{person}{Wayne~K. Goodman}, {and} \bibinfo{person}{Amy~L. McGuire}.} \bibinfo{year}{2018}\natexlab{}.
\newblock \showarticletitle{Continued access to investigational brain implants}.
\newblock \bibinfo{journal}{\emph{Nature Reviews Neuroscience}} \bibinfo{volume}{19}, \bibinfo{number}{6} (\bibinfo{date}{April} \bibinfo{year}{2018}), \bibinfo{pages}{317–318}.
\newblock
\showISSN{1471-0048}
\urldef\tempurl%
\url{https://doi.org/10.1038/s41583-018-0004-5}
\showDOI{\tempurl}


\bibitem[Rossi et~al\mbox{.}(2017)]%
        {BaitAndSwitch2017}
\bibfield{author}{\bibinfo{person}{P.~Justin Rossi}, \bibinfo{person}{James Giordano}, {and} \bibinfo{person}{Michael~S. Okun}.} \bibinfo{year}{2017}\natexlab{}.
\newblock \showarticletitle{The Problem of Funding Off-label Deep Brain Stimulation: Bait-and-Switch Tactics and the Need for Policy Reform}.
\newblock \bibinfo{journal}{\emph{JAMA Neurology}} \bibinfo{volume}{74}, \bibinfo{number}{1} (\bibinfo{date}{Jan.} \bibinfo{year}{2017}), \bibinfo{pages}{9}.
\newblock
\showISSN{2168-6149}
\urldef\tempurl%
\url{https://doi.org/10.1001/jamaneurol.2016.2530}
\showDOI{\tempurl}


\bibitem[Sriram et~al\mbox{.}(2023)]%
        {Sriram2023}
\bibfield{author}{\bibinfo{person}{Karthik Sriram}, \bibinfo{person}{Raghavendra~Pradyumna Pothukuchi}, \bibinfo{person}{Michał Gerasimiuk}, \bibinfo{person}{Muhammed Ugur}, \bibinfo{person}{Oliver Ye}, \bibinfo{person}{Rajit Manohar}, \bibinfo{person}{Anurag Khandelwal}, {and} \bibinfo{person}{Abhishek Bhattacharjee}.} \bibinfo{year}{2023}\natexlab{}.
\newblock \showarticletitle{SCALO: An Accelerator-Rich Distributed System for Scalable Brain-Computer Interfacing}. In \bibinfo{booktitle}{\emph{Proceedings of the 50th Annual International Symposium on Computer Architecture}} \emph{(\bibinfo{series}{ISCA ’23})}. \bibinfo{publisher}{ACM}.
\newblock
\urldef\tempurl%
\url{https://doi.org/10.1145/3579371.3589107}
\showDOI{\tempurl}


\bibitem[Stevenson and Kording(2011)]%
        {Stevenson2011}
\bibfield{author}{\bibinfo{person}{Ian~H Stevenson} {and} \bibinfo{person}{Konrad~P Kording}.} \bibinfo{year}{2011}\natexlab{}.
\newblock \showarticletitle{How advances in neural recording affect data analysis}.
\newblock \bibinfo{journal}{\emph{Nature Neuroscience}} \bibinfo{volume}{14}, \bibinfo{number}{2} (\bibinfo{date}{Jan.} \bibinfo{year}{2011}), \bibinfo{pages}{139--142}.
\newblock
\urldef\tempurl%
\url{https://doi.org/10.1038/nn.2731}
\showDOI{\tempurl}


\bibitem[Ugur et~al\mbox{.}(2024)]%
        {Swapping2024}
\bibfield{author}{\bibinfo{person}{Muhammed Ugur}, \bibinfo{person}{Raghavendra~Pradyumna Pothukuchi}, {and} \bibinfo{person}{Abhishek Bhattacharjee}.} \bibinfo{year}{2024}\natexlab{}.
\newblock \showarticletitle{Swapping-Centric Neural Recording Systems}. In \bibinfo{booktitle}{\emph{The 15th Annual Non-Volatile Memories Workshop}}.
\newblock


\end{thebibliography}

\end{document}